\documentclass[a4paper,12pt,draft]{article}
\newcommand{\dfrac}{\displaystyle \frac }

\newcommand{\nkg}[1]{\mbox {\boldmath $#1$}}
\newcommand{\slin}[1]{\mbox {\fontsize{6pt}{1mm}\selectfont $ #1 $}}% - позволяет написать индекс нужного размера сложной формы.

\title{
\Large{The Watch Paradox: Solution of the Problem} }
\author{I.A.Solomeshch}
\date{}
\begin{document}

\maketitle

\begin{abstract}
{\small The article presents the detailed analysis of the watch paradox. It is shown that it arose because of unjustified, as it turned out, identification of watch readings at the moment of its return with the time read  by it.}
\bigskip 

\end{abstract}
\section {Introduction}
\textbf{1.1}\quad   In his first article from 1905 [1] on the special theory of relativity (SR)  for inertial  reference systems( IRS ) Einstein made an unusual conclusion from Lorenz's transformations about the slowing of the moving watch; let's call it a lemma for reference. Basing on this lemma, Einstein  proved  an astonishing Statement 1 in the same article.

	\textit{{\large{Statement 1\footnote{The formulae and wordings are given regarding the one-dimensional spatial case, which we focus on.}.}}\quad In a certain point $a$ on the axis $x$ of some IRS $S$ let there be two similar synchronized watches $C$ and $C^\prime$. If you move one of them -- $C^\prime$ along the axis $x$ with the constant speed $v$ to some other point $b$ of the axis $x$, and then immediately turn-around and return it to the initial point with the same speed, then on returning, the reading of the moving watch will be less than that of the watch constantly  kept in point $a$.}\\
Here, the increment of the moving watch readings from the moment of motion start up to its return to the point $a$ will be
$$\gamma=\left(1-\frac{v^2}{c^2}\right)^{-1/2} \eqno(1.1)$$
times less than the increment readings of static watch in this interval ($c$ - is the vacuum light speed).

	The paradoxicality of Statement 1 is greatly increased if the moving watch is connected with the coordinate system with the axis $x^\prime$ directed along the axis $ x$ and the same unit of scale. Issuing from the relativity of motion, the watch $C^\prime$ can be considered static, and the watch $C$ -- relatively moving. Then, as it seems, Statement 1 with the role-change of watches $C$ and $C^\prime$ leads to Statement~2.
	
\textit{{\large{Statement 2.}}\quad 	 By the moment of return, it is the reading of the watch $C$ that will be less },\\
which, as it was considered, contradicts the former statement. This contradiction was called the watch paradox (the time, the twins paradox). So,

	\textit{{\large{The watch paradox}}\quad  consists in the fact that the proved Statements 1 and 2 seem incompatible.}
	
	The truth of the watch paradox, as it seemed, leads necessarily to the conclusion about the inner contradictions of SR.
	
	The further development of events can be given just in general.
	
	In 1911 Langevin [2] reformulated the statement about the slowing of the watch, moving along a closed route (Statement 1), into a statement about the age slowing of space travelers and the possibility of the farthest possible space travels during the cosmonaut's lifetime. In the same work Langevin pointed out Reason 1 against Statement 2.
	
\textit{{\large{Reason 1.}}\quad	The  reference system, connected with the moving watch  $C^\prime$, isn't inertial because of the acceleration that it inevitably undergoes due to the change in the direction of the movement, and the conclusions of SR are inapplicable to it\footnote{ This objection can also be referred to the substantiation of Statement 1 by Einstein.}.}\\
 So Statement 2 isn't valid, and the paradox is solved.

	In return Laue ([3] from 1913) noted that the acceleration is insignificant regarding the slowing of the moving watch on its return to the initial point, because its slowing could be made voluntarily great comparing with the change of watch readings at the sections of acceleration due to the increase of sections with even motion. The most important thing is that the watch $C^\prime$ is successively situated in two different IRSs. Besides, in [4] from 1913 Laue substantiated Statement 1 for a more real case, when the change of the watch $C^\prime$ speed goes evenly at the beginning, at the turn in the distant point and at the end of the route.
	
	However, Laue's argument can be used also in case, when it is the watch  $C$ that is considered moving, and this leads to Reason 2.
	
	\textit{{\large{Reason 2.}}\quad At the section of $C^\prime$ distancing from $C$ with constant  speed, as well as at the section of  returning to $C$  with constant speed, the frame of reference, connected with $C^\prime$, will be inertial, although different, for each of them. Then, at each of these sections, it is the watch $C$ that can be considered moving and, according to the lemma, it is it that will slow down. Hence, with the sufficient distancing of the watch $C$ from the watch $C^\prime$  turn-point and maintaining the acceleration regime at the turn, Statement 2 holds true and the paradox remains.}\footnote{I don't know, who and when first came to such substantiation of Statement 2. (It "must" have been pointed out by Laue.) Nevertheless, this substantiation is given and "is refuted" at once in [5, ch.3, \S 1] and in [6, ch.1, \S 2].} \\
	Moreover, it is possible with any given relative precision to provide the equality of the quotient of increments of watches readings, now $C^\prime$ to the increment of the watch $C$ readings, to Quantity 1.1.
	
	\textbf{1.2}\quad	Because of the great theoretical (for SR) and "practical" (after Lan-gevin's work [2]) importance, the simplicity of the wording and mathematical formula, the watch paradox drew attention of many professionals and amateurs.
	
"There is abundance of literature on the subject", wrote Marder back in 1971 in the
introduction to the book [5], more than 300  references, which was a detailed review of works on the watch paradox  for that year.

	Publications on this subject arrive in plenty (the latest known to the author work is [7] from 2007).
	
	To characterize this abundance of literature, let's consider the possible ways of the paradox solving.
	
	As it was recognized that Statements 1 and 2 are mutually contradicting, and that the proof of Statement 1 by means of SR causes no doubts, while the only known way of Statement 2 substantiation is Reason 2, there are only two possibilities:
	
\textit{{\large{Possibility 1.}}\quad Statement 2 is false.}

\textit{{\large{Possibility 2.}}\quad SR is contradictory.}

Hence, there are only three ways to solve the watch paradox.

\textit{{\large{Way 1.}}\quad To prove that Statement 2 cannot be proved by means of SR, and the paradox would disappear for ever (within the limits of SR).}

\textit{{\large{Way 2.}}\quad To show that the proof of Statement 2 given in Reason 2 is groundless. With this the paradox would disappear, at least, for  the present.}

\textit{{\large{Way 3.}}\quad To prove that SR is contradictory. This would destroy both the paradox and SR. }

	Now let's turn to the question of provability of attempts to solve the watch paradox.
	
	The first way hasn't yet been investigated by anybody \footnote{Here and below such statements presuppose the reservation "as far as the author knows".}. 
There are also no well substantiated  works investigating the third way. 

	All the works on the watch paradox finally tend to refute Statement 2  and fall into two groups according to the way of implementation of this task. 
	
	The first group includes the works repeating Langevin's argumentation (Reason 1) that by all means refutes all substantiations of Statement 2, of the type given before (before the wording of this statement), but hereby Reason 2 is not taken into consideration. In particular, here belong works [8 - 17].
	
	The second group includes works in this or that way proving Statement 1 and with the tacit supposing of SR truthfulness, Statement 2 is considered refuted. The value of these works on the paradox is only in the fact that they give different variants of Statement 1 proof. However, because Statement 2 presupposes a contradictoriness of SR, these works don't refute Statement 2, they simply consider it false at first. Here belong works [6; 18-21].
	
	Hence, the second way hasn't also been investigated by anyone, and it is impossible to disagree with the opinion stated in work [22]: "Ninety years, hundreds of books, thousands of articles, but this matter  still stands at the point von Laue left it."
	
\textbf{1.3}\quad 	This work shows that besides possibilities 1 and 2 there is yet a third one - Statement 2 also holds true and doesn't contradict Statement 1.  It became possible to demonstrate that the fact that the watch $C^\prime$ readings at its meeting with the watch $C$ are greater, doesn't yet mean that $C^\prime$ has read more time. The solving of the twin paradox proved to be found quite another way; it was only necessary to precise the contents of Statements 1 and 2. 

This situation is quite similar to the one happened to a Mr.N.

At 9 o'clock in the evening Mr.N started from the town clock moving to point A, intending to return by 6 o'clock in the morning next day. Last week Mr.N had already completed the same route during 9 hours. Mr.N traveled the distance keeping the speed of his previous trip exactly. However, when Mr.N returned, he was struck - the town clock read 7 o'clock. This mysterious phenomenon could be easily explained: Mr.N neglected the fact that the previous night the town clock  hands had been moved one hour forward because of the start of the summer time. 
\section{ Additional information. Setting the task}

	This work considers only the one-dimensional spatial IRSs $S, S^\prime$ etc. of the special theory of relativity, with  the coordinate axes $x, x^\prime$ etc. directed the same way and sliding one on another with a relative movement. The points of the coordinate axis $x$ ($x^\prime$ etc.) have watches synchronized among themselves according to Einstein. Let's name $t$ ($t^\prime$ etc.) the time of the IRS $S$ ($S^\prime$ etc.) and $O$ ($O^\prime$ etc.) - the origin  of coordinates on the axis $x$ ($x^\prime$ etc).
	
	\textbf{2.1 Additional information.} \quad  For convenience let's cite the following most frequently used information from SR.
	 
	\textit{2.1.1} \quad Let $\nkg A$  be a point event, i.e. an event taken place at a certain point of space, at a certain point of time. 
	
$x_{\slin {\nkg A}}$ is the coordinate of the point on the axes $x$, where the event $\nkg A$ took place, and $t_{\slin {\nkg A}}$ - the time according to the watch placed in that point, when that event took place, are called space-time coordinates  of $\nkg A$ in the IRS $S$. In a word, they speak about the event $\nkg A$ taking place in the space-time point ($x_{\slin {\nkg A}}, t_{\slin {\nkg A}}$) of the  system $S$.

	If we know the space-time coordinates of some event $\nkg A$ in the system $S$ - ($x_{\slin {\nkg A}}, t_{\slin {\nkg A}}$), and in the system $S^\prime$ - ($x_{\slin {\nkg A}}^\prime,  t_{\slin {\nkg A}}^\prime$) (i.e. this event took place in the point $x_{\slin {\nkg A}}$ of the axis  $x$ at the moment, when this point coincided with the point $x_{\slin {\nkg A}}^\prime$  of the axis $x^\prime$, and the watches of the systems $S$ and $S^\prime$ placed in this point read the time $t_{\slin {\nkg A}}$ and $t_{\slin {\nkg A}}^\prime$ accordingly), the space-time coordinates of the IRSs  $S^\prime$ and $S$ are connected by Lorenz's transformation 
							$$
\left\{
\begin{array}{l}
x^\prime - x^\prime_{\slin {\nkg A}} = \gamma \left\{(x - x_{\slin {\nkg A}}) - v(t - t_{\slin {\nkg A}})\right\}\\
t^\prime - t^\prime_{\slin {\nkg A}} = \gamma \left\{(t - t_{\slin {\nkg A}}) - \dfrac{v}{c^2}(x - x_{\slin {\nkg A}})\right\},\quad  \forall x, t	,
\end{array}
\right. \eqno(2.1)
$$
where $v$ is the speed of $S^\prime$ relatively to $S$, and {$\gamma $ is defined by (1.1).

Let's call the event $\nkg A$ coordinating the  systems $S$ and $S^\prime$.

The literature on the subject cites more often the special case, when
		$$(x_{\slin {\nkg A}}, t_{\slin {\nkg A}}) = (x_{\slin {\nkg A}}^\prime, t_{\slin {\nkg A}}^\prime) = (0, 0)\;,$$
i.e. when the origins of coordinates $O$ and $O^\prime$ coincide just when the watches of both systems, placed in these points, read the same time - zero:
					$$
\left\{
\begin{array}{l}
x^\prime  = \gamma \left( x  - vt \right)\\
t^\prime  = \gamma \left(t  - \dfrac{v}{c^2}x \right),\quad \forall x, t .	
\end{array}
\right. \eqno(2.2)
$$

The system of equations (2.2) connects four variables  $x, t, x^\prime, t^\prime $, while the two latter ones are expressed in it by means of the two former ones. Choosing an arbitrary pair of variables, basing on (2.2), the rest can be expressed through them.

Let's consider the variants that will be necessary further. 
$$
\left\{
\begin{array}{l}
x  =  \dfrac {x^\prime}{\gamma}   + vt \\
t^\prime  = \dfrac {t}{\gamma}  - \dfrac{v}{c^2}x^\prime \;,
\end{array}
\right. \eqno(2.3)
$$

$$ 
\left\{
\begin{array}{l}
t= \dfrac {1}{v} \left( x - \dfrac {x^\prime}{\gamma} \right )\\
t^\prime = \dfrac {1}{v} \left( \dfrac {x}{\gamma} - x^\prime \right),\quad v \neq 0\;,	
\end{array}
\right.  \eqno (2.4)
$$

$$
\left\{
\begin{array}{l}
x^\prime  =  \dfrac {x}{\gamma}   - vt^\prime \\
 t = \dfrac {t^\prime}{\gamma}  - \dfrac{v}{c^2}x \;.
\end{array}
\right. \eqno(2.5)
$$

The analogous variants for the system (2.1) are got from (2.3)-(2.5) through the change of $x, t, x^\prime, t^\prime $ into $x-x_{\slin {\nkg A}}, t-t_{\slin {\nkg A}} , x^\prime - x^\prime_{\slin {\nkg A}}, t^\prime - t^\prime_{\slin {\nkg A}} $ accordingly; we'll refer to them as to $(2.3_{\slin {\nkg A}})$ -- $(2.5_{\slin {\nkg A}})$. If the coordinating event is named with another letter, let's substitute it for $\nkg A $.

\textit{2.1.2}\quad Let $S^\prime$ and $S$ be arbitrary IRSs,  $v$ - the speed of $S^\prime$ relatively to~ $S$.

\textit{{\large {Lemma } (about the slowing of the moving watch)}}

\textit{If the watch fixed at a certain point of the axis $x^\prime$ at the moment $t^\prime _1$ read by this watch, passed by the watch fixed on the axis $x$ and reading time $t_1$, and then at the moment $t^\prime _2$ passed by the watch fixed on the axis $x$ and reading  time $t _2$, then}
			$$ t_2^\prime - t_1^\prime = \frac{1}{\gamma} (t_2 - t_1).$$
			
In other words, the time interval between two events taking place at the fixed point of system $S^\prime$, read  by the watch of this system situated in this point, is $\gamma$ times less the time interval calculated according to the readings of the watches, static in the system $S$ and placed in the points where these events took place.

\textbf{2.2 Setting the task } \quad    Studying the watch paradox, we adopt the scheme of the instant speed $C^\prime$ change into the opposite. Of course, the instant turn back of a real physical object is impossible, but here we speak about the model task within the mathematical model of space-time relations  between different IRSs in SR.

The instant change of speed direction preserving its module  was used by Einstein in [1] from 1905 and, much later after "the discussion Langevin - Laue", by other famous specialists in SR (see, for example, [12, 13, 19, 20]). Taking also into consideration that the watch paradox arose just while using the scheme of "the instant turn" and that the results of calculations according to this scheme enable  to get a very good approximation for the case of even speed changing (substantiations by Laue [3, 4] and similar to them in Reason 2 and up to the  point 1.2, then the adoption of this scheme is justified.

Besides, we can admit that there is no acceleration and slowing of the watch $C^\prime$ near the watch $C$, considering that it passes by $C$ with the speed $v$ at first, and in the end -- with the speed $-v$.

\textit{{\large Task.}}\\
\textit{Let's name $C$ ($C^\prime$) the watch of the system $S$ ($S^\prime$) placed at the origin of coordinates on the axis $x$ ($x^\prime$). Let:}

\textit{1)	The IRS $S^\prime$ move relatively to the IRS $S$ with a relative speed $v>0$;}

\textit{2)	The watch $C^\prime$ moves by $C$ (the first meeting - the event $\nkg B$) at the moment, when the watches $C^\prime$ and $C$ read the time $t^\prime = 0$ and $t=0$ accordingly;}

\textit{3)	In $t_{\slin {\nkg R }}$ time by the watch  of the system $S$ after the meeting of watches $C^\prime$ and $C$, the watch $C^\prime$ instantly changes its speed from $v$ to $-v$ (the event $\nkg R$), still remaining static relatively to the axis $x^\prime$. Thus, in the process of reverse movement the watch $C^\prime$ will remain at the origin of coordinates of the same axis $x^\prime$, but in another IRS.
\footnote{ Such like operation is considered in [1,\S3].}.}

\textit{Taking into consideration the relativity of movement, it is necessary to calculate the readings of the watches $C^{\prime}$ and $C$ at the moment of their second meeting, i.e. when the watches $C^{\prime}$ and $C$ are found at the same level (the event $\nkg E$),}

\textit{a) considering the system $S$ and the watch $C$ static (task 1) and}

\textit{b) considering the systems $S^\prime$ and the new one in succession and the watch} 

 \textit{$C^\prime$ static (task~2).}
\section{ Solving of tasks 1 and 2 }
The solving of tasks 1, 2 is of the same type. The process of changing of the watches $C$ and $C^\prime$ mutual position between the events, the first and the second meetings are subdivided into two stages: the first meeting -- the turn, and the turn -- the second meeting. Then for each of the events, marking these stages, the readings of the moving watch ($C^\prime$ or $C$, depending on the viewpoint) are studied at the moment of this event, as well as the readings of the watch of  the "static" IRS at the same moment at the point, where the event took place.

	We suggest two ways of solving each task: the simplest, but nevertheless strict, and more formal, but enabling to analyze the result in details.
	
\textbf {3.1  Solving of task 1}

\textit{3.1.1. Simplest solving.}

\textit{Stage $\nkg B$ -- $\nkg R$}.\quad According to  term 2) of the task  the watch $C^\prime$ is in the point $x=0$ of  the system $S$ at the moment $t=0$, i.e. at the moment of the event $\nkg B$. Because $C^\prime$  moves relatively to $S$ with the speed $v$, then in the time interval $t_{\slin{\nkg R}}$ (i.e. by the moment of the turn, i.e. - the event $\nkg R$) it will be found in the point $x_{\slin{\nkg R}} = v t_{\slin{\nkg R}}$ just when the watch of the system $S$ reads the time $t_{\slin{\nkg R}}$ in this point.

	According to the lemma about the slowing of  the moving watch,
$$ t^\prime_{\slin{\nkg R}} - t^\prime_{\slin{\nkg B}} = \dfrac{1}{\gamma} \  (t_{\slin{\nkg R}} - t_{\slin{\nkg B}}), $$
considering that according to the term 2) $t^\prime_{\slin{\nkg B}} = t_{\slin {\nkg B}} = 0,$
$$ t^\prime_{\slin{\nkg R}} = \dfrac{1}{\gamma} \ t_{\slin{\nkg R}}. \eqno (3.1)$$

Remember that the events $\nkg B$ and $\nkg R$ take place in the point of the watch $C^\prime$ position, and that is why $t^\prime_{\slin{\nkg B}}$ and $t^\prime_{\slin{\nkg R}}$ are the readings of the watch $C^\prime$ at the moment of these events.

\textit{Stage $\nkg R$ -- $\nkg E$}. \quad 	At the point $x_{\slin{\nkg R}}\;$  $C^\prime$ instantly changes its speed into $-v$, still remaining static relatively to the axis $x^\prime$. So in the process of reverse movement the watch $C^\prime$ will be situated at the origin of coordinates of the same axis $x^\prime$, but in another IRS.  Let's name it $S^{\prime \prime}$ and rename the axis $x^\prime$ and the watch $C^\prime$ after the turn into $x^{\prime \prime}$ and $C^{\prime \prime}$ to avoid ambiguity.
 
	Readings of the watch $C^\prime$ at the moment of the  turn don't change, so we have
			$$t^{\prime \prime}_{\slin{\nkg R}}=t^{\prime }_{\slin{\nkg R}}. \eqno (3.2)$$
			
	Because at the moment $t_{\slin{\nkg R}}\;$  $C^{\prime \prime}$ was found at the point $x = x_{\slin{\nkg R}}>0 $, then, moving with the speed $-v$ relatively to the axis $x$ it will be found at the point $x=0$ at the moment $ t_{\slin{\nkg R}} + \dfrac{x_{\slin{\nkg R}}}{v} = 2t_{\slin{\nkg R}}.$ So at the moment of the event $\nkg E $ the watch $C$ reads the time
			$$ t_{\slin{\nkg E}} =  2t_{\slin{\nkg R}}. \eqno (3.3)$$

	According to the lemma $t^{\prime \prime}_{\slin{\nkg E}} - t^{\prime \prime}_{\slin{\nkg R}} = \dfrac{1}{\gamma} (t_{\slin{\nkg E}} - t_{\slin{\nkg R}}).$ Taking into consideration (3.2), (3.1) and (3.3), we get $t^{\prime \prime}_{\slin{\nkg E}} = \dfrac{2t_{\slin{\nkg R}}}{\gamma}$.
	
	So the watches $C$ and $C^{\prime \prime}$ (that is $C^{\prime}$) at the moment of their second meeting (the event $\nkg E$) read:
$$t_{\slin{\nkg E}} = 2t_{\slin{\nkg R}}, \quad  t^{\prime \prime}_{\slin{\nkg E}} = \dfrac{2t_{\slin{\nkg R}}}{\gamma}, \quad  \dfrac{t^{\prime \prime}_{\slin{\nkg E}}}{t_{\slin{\nkg E}}} = \dfrac{1}{\gamma}. \eqno (3.4)   $$

\textit{3.1.2. More formal solving}

\textit{Stage $\nkg B - \nkg R.$} \quad   According to term 2) of the task, the coordinates of the event $\nkg B $ in the system $S$ are $(x_{\slin{\nkg B}}, \,t_{\slin{\nkg B}}) = (0,0)$, and in the system $S^\prime$ --- $\;(x^\prime_{\slin{\nkg B}}, \,t^\prime_{\slin{\nkg B}}) = (0,0)$.

	Considering the event $\nkg B $ as coordinatig  for $S$ and $S^\prime$, we see that the relation between the coordinates of  the events is presented in these systems with  any of the equation systems (2.2) - (2.5).
	
	According to the very sense of the event $\nkg R$, it is known that it takes place at the point $x^\prime = 0 $ at the moment $t = t_{\slin{\nkg R}}$ by the watch of the  system $S$. The two other coordinates of $\nkg R$ --- $\;x$ and $t^\prime$ can be easily found from (2.3), substituting the already known $x^\prime$ and $t$:
		$$x =  \left. \left (\dfrac {x^\prime}{\gamma} + vt \right)\right|_{\slin{\nkg R}} = vt_{\slin{\nkg R}} $$
$$t^\prime = \left.\left (\dfrac {t}{\gamma} -  \dfrac {v}{c^2} x^\prime \right)\right|_{\slin{\nkg R}} = \dfrac {t_{\slin{\nkg R}}}{\gamma}. $$
So the coordinates of $\nkg R$ in the systems $S$ and $S^\prime$ are 
		$$(x_{\slin{\nkg R}},t_{\slin{\nkg R}}) = (vt_{\slin{\nkg R}},t_{\slin{\nkg R}}), \; (x^\prime_{\slin{\nkg R}},t^\prime_{\slin{\nkg R}}) = (0 ,t_{\slin{\nkg R}}/\gamma) \eqno (3.5)  $$
accordingly. But the coordinates of the event $\nkg R$  in the systems $S^\prime$ and $S^{\prime \prime}$  are the same (it's enough to recall that this event takes place at the origin of coordinates of both systems, and consider (3.2), so we have
			$$(x^{\prime\prime}_{\slin{\nkg R}},t^{\prime\prime}_{\slin{\nkg R}}) = (0 ,t_{\slin{\nkg R}}/\gamma). \eqno (3.6) $$
			
\textit{Stage $\nkg R - \nkg E.$} \quad 	  Considering  $\nkg R$ as an event, coordinating the systems $S$ and $S^{\prime \prime}$, we arrive at the conclusion that the coordinates of events in them are related through the system of equations (2.1) and the systems $(2.2_{\slin {\nkg R}}) - (2.5_{\slin {\nkg R}} )$  equivalent to it (see the last passage of point (2.1.1) after the substitution  of $v$ into $-v$ in all these systems of equations (because the speed of $S^{\prime \prime}$ relatively to $S$ is $-v$) and $^\prime$ into $^{\prime \prime}$. For the calculation of $t_{\slin{\nkg E}}$ and $t^{\prime\prime}_{\slin{\nkg E}}$  that are of interest to us, because according to the very wording of the event $\nkg E$ it is known that $x_{\slin{\nkg E}} =  x^{\prime \prime} _{\slin{\nkg E}} = 0$, it is convenient to make use of the system $(2.4_{\slin {\nkg R}})$. Considering (3.5), (3.6), we get
	$$t - t_{\slin{\nkg R}} = -\dfrac{1}{v} \left. \left\{(x-x_{\slin{\nkg R}}) - \dfrac{x^{\prime\prime} - x^{\prime\prime}_{\slin{\nkg R}}}{\gamma} \right\}\right|_{\slin{\nkg E}} = \dfrac {x_{\slin{\nkg R}}}{v} - \dfrac{x^{\prime\prime}_{\slin{\nkg R}}}{v\gamma},  $$ 
$$t^{\prime\prime} - t^{\prime\prime}_{\slin{\nkg R}} = -\dfrac{1}{v} \left. \left\{\dfrac{x-x_{\slin{\nkg R}}}{\gamma} - (x^{\prime\prime} - x^{\prime\prime}_{\slin{\nkg R}}) \right\}\right|_{\slin{\nkg E}} = \dfrac {x_{\slin{\nkg R}}}{v\gamma} - \dfrac{x^{\prime\prime}_{\slin{\nkg R}}}{v}  $$
and then
$$ t =  t_{\slin{\nkg R}} + \dfrac {x_{\slin{\nkg R}}}{v} = 2t_{\slin{\nkg R}},\; t^{\prime\prime} = t^{\prime\prime}_{\slin{\nkg R}} + \dfrac {x_{\slin{\nkg R}}}{v\gamma} = \dfrac{2t_{\slin{\nkg R}}}{\gamma}, $$ i.e.
$$ t_{\slin{\nkg E}} = 2t_{\slin{\nkg R}}\,, \;\; t^{\prime\prime}_{\slin{\nkg E}} = \dfrac {2t_{\slin{\nkg R}}}{\gamma}\,,\;\; \dfrac{t^{\prime\prime}_{\slin{\nkg E}}}{t_{\slin{\nkg E}}} = \dfrac {1}{\gamma}\,, \eqno (3.7)$$ that coincides with the result arrived at before (see (3.4)).

\textbf{3.2. Solving of task 2}

Now it is the watch $C$ that should be considered moving together with the IRS $S$.  At first let's study  the movement of $S^\prime$ relatively to $C$ and only then let's reformulate the things, considering the case, when it is the watch $C$ that is considered moving.

At the moment $t^\prime=0$ the watch $C^\prime$, situated in the point $O^\prime$, passes by $C$, situated in the point $O$, to the right, or to be more precise, in the positive direction of the axis $x$ , with the speed $v$. After some yet unknown time $t^\prime_{\slin{\nkg L}}$, when the watch $C$ reaches the watch  $C_1^\prime$ of the system $S^\prime$, situated in the  yet unknown point $x^\prime = x^\prime_{\slin{\nkg L}} $, and reading the time $t^\prime = t^\prime_{\slin{\nkg L}}$, the watch $C_1^\prime$ instantly changes its speed relatively to the axis $x$ into $-v$ (the turn -- the event $\nkg L$,\footnote{$x^\prime_{\slin{\nkg L}}$ and $t^\prime_{\slin{\nkg L}}$ are the coordinates of the event $\nkg L$ in the system $S^\prime$, so the agreement about the coordinates of events names (see 2.1.1.) is not broken here.} ) and without changing its position on the axis $x^\prime$, it starts moving  to the left along the axis $x$.
 
In the process of reverse movement the watch $C_1^\prime$ will still remain in the point $x^\prime = x^\prime_{\slin{\nkg L}}$ of the axis $x^\prime$, but in another IRS. Let's name it $S^{\prime\prime\prime}$, and  rename the axis $x^\prime$, the watch $C_1^\prime$ and $C^\prime$ after the turn into $x^{\prime\prime\prime} $, $C_1^{\prime \prime \prime}$, $C^{\prime \prime \prime}$ accordingly. The readings of the watch $C_1^\prime$ at the moment of the turn don't change, so we have
	$$t^{\prime\prime\prime}_{\slin{\nkg L}} = t^\prime_{\slin{\nkg L}}.  \eqno (3.8)$$
	
Let's dwell upon the solving of task 2 and consider the watch $C$ moving.
\textit{3.2.1. Simplest solving}

\textit{Stage}  $\nkg B$ -- $\nkg L$. \quad  The watch $C$ moves  relatively to $x^\prime$  with the speed $-v$ and at the moment $t^\prime = 0$ is in the point  $x^\prime = 0$ (the event  $\nkg B$). Hence, after the time interval $t^\prime_{\slin{\nkg L}}$, i.e. by the moment of the turn (the event $\nkg L$) it will be found at the point $x^\prime_{\slin{\nkg L}} = -vt^\prime_{\slin{\nkg L}}$, when the watch $C_1^\prime$ of the system $S^\prime$, situated in this point, reads the time $t^\prime_{\slin{\nkg L}}$.
	According to the lemma, it is $C$ that is moving now,
			$$t_{\slin{\nkg L}} - t_{\slin{\nkg B}} = \dfrac{1}{\gamma} \ (t^\prime_{\slin{\nkg L}} - t^\prime_{\slin{\nkg B}}).$$
Taking into consideration that $t^\prime_{\slin{\nkg B}} = t_{\slin{\nkg B}} = 0 $, we have
			$$t_{\slin{\nkg L}} = \dfrac{1}{\gamma} \  t^\prime_{\slin{\nkg L}}. \eqno (3.9)$$
			
Because the events $\nkg B$ and $\nkg L$ take place in the point, where the watch $C$ is situated, then $t_{\slin{\nkg B}}$ and $t_{\slin{\nkg L}}$ are the readings of the watch $C$ at the moment of these events.

	\textit{Stage }$\nkg L$ -- $\nkg E$. \quad At the moment $t^\prime_{\slin{\nkg L}} = t^{\prime\prime\prime}_{\slin{\nkg L}}$ (see (3.8)) the speed of the watch $C$ relatively to the axis $x^{\prime\prime\prime}$  (the former $x^\prime$) instantly gets equal to $v$. Because at the moment $t^{\prime\prime\prime}_{\slin{\nkg L}} \; $    the watch $C$ was found in the point $x^{\prime\prime\prime} =  x^{\prime\prime\prime}_{\slin{\nkg L}} = x^\prime_{\slin{\nkg L}}$, then at the moment $t^{\prime\prime\prime} = t^{\prime\prime\prime}_{\slin{\nkg L}} + |x^{\prime\prime\prime}_{\slin{\nkg L}}|/v = 2t^{\prime\prime\prime}_{\slin{\nkg L}}$ it will be found in $O^{\prime\prime\prime}$, i.e. it will reach the watch $C^{\prime\prime\prime} = C^\prime$  (the event $\nkg E$)\footnote{ The coordinate of the point on the axis $x^{\prime\prime\prime}$, where the watch $C^{\prime\prime\prime}$ (that is $C^\prime$) is fixed, is still zero, i.e. it is in the origin of coordinates in $S^{\prime\prime\prime}$ also. It is because the distance between the points marked on the line, while preserving the scale, doesn't depend on the speed of the even rectilinear movement of this line relatively to some IRS ([23, \S 5], [13, ch. 6, \S 5]).}.
	So $t^{\prime\prime\prime}_{\slin{\nkg E}} = 2t^{\prime\prime\prime}_{\slin{\nkg L}}$, and because according to the lemma $t_{\slin{\nkg E}} - t_{\slin{\nkg L}} = \dfrac{1}{\gamma} \ (t^{\prime\prime\prime}_{\slin{\nkg E}} - t^{\prime\prime\prime}_{\slin{\nkg L}}),$ then, taking into consideration (3.9) and (3.8), we arrive at $t_{\slin{\nkg E}} = 2t^\prime_{\slin{\nkg L}}/\gamma$.
	
	 So the watches $C$ and $C^{\prime \prime\prime}$ (that is $C^{\prime}$) at the moment of their second meeting (the event $\nkg E$) read:
   $$ t_{\slin{\nkg E}} = \dfrac{2t^\prime_{\slin{\nkg L}}}{\gamma} \; , t^{\prime\prime\prime}_{\slin{\nkg E}} = 2t^\prime_{\slin{\nkg L}} \; , t^{\prime\prime\prime}_{\slin{\nkg E}} / t_{\slin{\nkg E}} = \gamma . \eqno (3.10)$$

\textit{3.2.2 More formal solving}

\textit{	Stage } $\nkg B$ -- $\nkg L$. Like in 3.1.2 the relation between coordinates of events in $S$ and $S^\prime$ is given with any of the equation systems (2.2) - (2.5).

	The event $\nkg L$ takes place in the point $x=0$ at the moment $t^\prime = t^\prime_{\slin{\nkg L}}$ by the watch of the system $S^\prime$. The two other coordinates of $\nkg L$ in the systems $S $ and $S^\prime$ -- $t$ and $x^\prime$ - are easy to calculate according to (2.5):
$$ x^\prime = \left. \left(\dfrac{x}{\gamma}-vt^\prime\right)\right|_{\slin{\nkg L}} = -vt^\prime_{\slin{\nkg L}},\; t = \left.\left(\dfrac{t^\prime}{\gamma} + \dfrac{v}{c^2}x\right)\right|_{\slin{\nkg L}} = \dfrac{t^\prime_{\slin{\nkg L}}}{\gamma}.$$
So the coordinates of    $\nkg L$ in the systems  $S$ и  $S^\prime$  ---    
$$ (x_{\slin{\nkg L}},t_{\slin{\nkg L}}) = (0,t^\prime_{\slin{\nkg L}}/\gamma),\; (x^\prime_{\slin{\nkg L}},t^\prime_{\slin{\nkg L}}) = (-vt^\prime_{\slin{\nkg L}},t^\prime_{\slin{\nkg L}}). \eqno (3.11)$$

The coordinates of the event $\nkg L$ in the systems $S^\prime$ and $S^{\prime\prime\prime}$ are the same, hence
	$$(x^{\prime\prime\prime}_{\slin{\nkg L}},t^{\prime\prime\prime}_{\slin{\nkg L}}) = (-vt^\prime_{\slin{\nkg L}},t^\prime_{\slin{\nkg L}}). \eqno (3.12)$$  
	
\textit{Stage} $\nkg L$ -- $\nkg E$. Considering the event $\nkg L$ as coordinating the systems $S$ and $S^{\prime\prime\prime}$, we arrive at the conclusion that the coordinates of events in them are related with any system of equations ($2.2_{\slin{\nkg L}}$) --  ($2.5_{\slin{\nkg L}}$) after the substitution of $v$ into $-v$ and $^\prime$ into $^{\prime\prime\prime}$ in them.

	Because $x_{\slin{\nkg E}} = x^{\prime\prime\prime}_{\slin{\nkg E}} = 0$, the $t_{\slin{\nkg E}}$ and $t^{\prime\prime\prime}_{\slin{\nkg E}}$ sought can be easily calculated according to $(2.4_{\slin{\nkg L}})$. Taking into consideration (3.11), (3.12), we arrive at
		$$t - t_{\slin{\nkg L}} = - \dfrac{1}{v}\left.\left \{(x - x_{\slin{\nkg L}}) -  \dfrac{x^{\prime\prime\prime} - x^{\prime\prime\prime}_{\slin{\nkg L}}}{\gamma}\right \}\right |_{\slin{\nkg E}} = \dfrac{x_{\slin{\nkg L}}}{v} - \dfrac{x^{\prime\prime\prime}_{\slin{\nkg L}}}{\gamma v}\,,$$
 $$t^{\prime\prime\prime} - t^{\prime\prime\prime}_{\slin{\nkg L}} = - \dfrac{1}{v}\left.\left \{\dfrac{x - x_{\slin{\nkg L}}} {\gamma} -  (x^{\prime\prime\prime} - x^{\prime\prime\prime}_{\slin{\nkg L}})\right \}\right |_{\slin{\nkg E}} = \dfrac{x_{\slin{\nkg L}}}{\gamma v} - \dfrac{x^{\prime\prime\prime}_{\slin{\nkg L}}}{ v}$$
and then
		 $$t = t_{\slin{\nkg L}} - \dfrac{x^{\prime\prime\prime}_{\slin{\nkg L}}}{\gamma v} = \dfrac {2t^\prime_{\slin{\nkg L}}} {\gamma}\;,t^{\prime\prime\prime} = t^{\prime\prime\prime}_{\slin{\nkg L}} - \dfrac{x^{\prime\prime\prime}_{\slin{\nkg L}}}{ v} = 2t^\prime_{\slin{\nkg L}}\;,$$
i.e.
$$t_{\slin{\nkg E}} = \dfrac {2t^\prime_{\slin{\nkg L}}} {\gamma}\;, t^{\prime\prime\prime}_{\slin{\nkg E}}  = 2t^\prime_{\slin{\nkg L}}\;, \dfrac {t^{\prime\prime\prime}_{\slin{\nkg E}}} {t_{\slin{\nkg E}}} = \gamma\;, \eqno (3.13)$$
that coincides with (3.10).

Now let's study the relation between $t^\prime_{\slin{\nkg L}}$ and $t_{\slin{\nkg E}}$.

According to terms 3), 1) and 2) of the task, the time $t_{\slin{\nkg R}}$, when the event $\nkg R$ took place, is known, and 
$$x^\prime_{\slin{\nkg R}}=0, \; x_{\slin{\nkg R}} = vt_{\slin{\nkg R}}. \eqno (3.14)$$

For the calculation of $x^{\prime\prime\prime}_{\slin{\nkg R}}$, we make use of the first equation of the
system ($2.2_{\slin{\nkg L}}$) and (3.12), (3.14), (3.11):
$$ x^{\prime\prime\prime}_{\slin{\nkg R}}  = x^{\prime\prime\prime}_{\slin {\nkg L}} +  \gamma \left.\left\{(x - x_{\slin {\nkg L}}) + v(t - t_{\slin {\nkg L}})\right\}\right|_{\slin{\nkg R}} = -vt^\prime_{\slin {\nkg L}} + \gamma \left \{ vt_{\slin {\nkg R}} + v(t_{\slin {\nkg R}}- t^\prime_{\slin {\nkg L}}/\gamma)\right \} =$$ 
$$= 2v(\gamma t_{\slin {\nkg R}} - t^\prime_{\slin {\nkg L}} ). $$
 According to Note $^7$  $x^{\prime\prime\prime}_{\slin {\nkg R}} = 0$, hence
$$ t^\prime_{\slin {\nkg L}} = \gamma t_{\slin R}  \;. \eqno (3.15) $$

Now the correlation (3.10) $\equiv$ (3.13) for the watches $C$ and $C^\prime$ readings at the moment of their second meeting can be, like in (3.7), expressed through  $t_{\slin{\nkg R}}$:
		$$ t_{\slin{\nkg E}} = 2t_{\slin{\nkg R}} \; , t^{\prime\prime\prime}_{\slin{\nkg E}} = 2\gamma t_{\slin{\nkg R}} \; , t^{\prime\prime\prime}_{\slin{\nkg E}} / t_{\slin{\nkg E}} = \gamma . \eqno (3.16).$$
\section{	 Solving of the paradox}
Section 3 presented a detailed, with details in literature usualy only implied at best, solving of the task about the watches $C$ and $C^\prime$ readings at the moment of their second meeting.

	It turned out that the readings of the watch that is conventionally considered static will be $\gamma$ times bigger.
	
	Because the readings of the watches at the moment of their first meeting were the same - zero, the readings at the moment of their second meeting were identified in literature as time intervals, read by the watches between the first and the second meetings.
	
	This very fact brought the watch paradox, according to which it is not the gist of the phenomenon that determines which of the watches reads greater time, but the fact which of them is conventionally considered static.
	
	It is shown next that watch readings and the time read by it  in this task they are, generally speaking, different quantities, and regardless of the method of calculation, the longer time interval will be read by the watch, fixed in one and the same IRS all the time.
	
	The watch paradox can be explained naturally, it ceases to be a paradox.
	
Let's pass on to the analysis as to what is the relation of watch readings at the moment of their second meeting: $t_{\slin{\nkg E}},    t^{\prime\prime}_{\slin{\nkg E}}$ - in task 1 and $t_{\slin{\nkg E}}, t^{\prime\prime\prime}_{\slin{\nkg E}}$ - in task 2 - to the time interval between the first and the second meetings, read by these watches, which we'll name $\tau_{\slin{\nkg E}}, \tau^{\prime\prime}_{\slin{\nkg E}}$ and $\tau^{\prime\prime\prime}_{\slin{\nkg E}}$ accordingly. Remember that the meaning of $t_{\slin{\nkg E}}$ in the first and the second tasks is the same.

\textbf{4.1 In task 1}	 it is the  IRS $S$ that was considered static. The watch $C$ of this system, situated in the point $O$, worked all the time between the first meeting (the event $\nkg B$) and the second meeting (the event $\nkg E$) in its natural way, and that is why the difference in its readings $t_{\slin{\nkg E}} - t_{\slin{\nkg B}} = t_{\slin{\nkg E}}$ is the time, read by this watch between the events $\nkg B$ and $\nkg E$, i.e.
$$\tau_{\slin{\nkg E}} = t_{\slin{\nkg E}}.  \eqno(4.1)$$

All the time the watch $ C^\prime $ had been situated in the origin of coordinates of the axis $x^\prime$ (renamed into $x^{\prime\prime}$ after the turn). It was the watch in the system $S^{\prime}$ from the moment of the event $\nkg B$ till the moment of the turn (the event $\nkg R$).

	At the moment of the turn the watch $C^\prime$ was included in the IRS $S^{\prime\prime}$ and renamed into the watch  $C^{\prime\prime}$ , but its readings weren't changed (see (3.2)), and it continued to work in its natural way till the moment of the second meeting, but in the system $S^{\prime\prime}$ . That is why the time, read by the watch $C^\prime$ (later renamed into $C^{\prime\prime}$) between the events $\nkg B$ and $\nkg E$ is
	$$\tau_{\slin{\nkg E}}^{\prime\prime} = (t_{\slin{\nkg E}}^{\prime\prime} - t_{\slin{\nkg R}}^{\prime\prime}) + (t_{\slin{\nkg R}}^\prime - t_{\slin{\nkg B}}^\prime) \\=  t_{\slin{\nkg E}}^{\prime\prime} - t_{\slin{\nkg B}}^\prime = t_{\slin{\nkg E}}^{\prime\prime} \;. $$
Here the fact that $t_{\slin{\nkg B}}^\prime = 0\;$ and $t_{\slin{\nkg R}}^{\prime\prime} = t_{\slin{\nkg R}}^\prime \; $ has been taken into consideration.

So,
			$$\tau^{\prime\prime}_{\slin{\nkg E}} = t^{\prime\prime}_{\slin{\nkg E}} . \eqno (4.2)$$
			
So (see (4.1), (4.2)) in task 1 the readings of the watches $C$ and $C^\prime$ at the moment of their second meeting coincide with the time, read by them between the first and the second meetings, i.e.(see (3.7))
		$$\tau_{\slin{\nkg E}} = 2t_{\slin{\nkg R}} \;, \tau_{\slin{\nkg E}}^{\prime\prime} = 2t_{\slin{\nkg R}}/\gamma. \eqno (4.3)$$
		
	Let's pay attention to the fact that the readings of the watch $C^\prime$ were continuously changing during all the time period, because it was included in the system $S^{\prime\prime}$, keeping the readings at the moment of the turn. Because the readings of $C^\prime$ was set at the moment of the turn in the system $S^{\prime\prime}$, the remaining watch of the system $S^\prime$ after the transformation into the system $S^{\prime\prime}$ was to be synchronized according to the watch $C^\prime$, and a priori it's not clear if it has kept the continuity of readings at the moment of its movement direction change\footnote{In the point 4.2 it will become clear that it hasn't kept it.}.

\textbf{4.2 In task 2} it is the watch $C^\prime$ and the IRSs  $S^\prime$ and $S^{\prime\prime\prime}$, related to it in turn, that were considered static. As to the readings of the watch $C$, everything mentioned in the point 4.1 remains true, so (4.1), i.e. $\tau_{\slin{\nkg E}} = t_{\slin{\nkg E}}$
 still remains true.
  
	It is yet necessary to analyze the readings of the watch $C^{\prime\prime\prime}$ (the watch $C^\prime$ renamed) at the moment of its second meeting - $t_{\slin{\nkg E}}^{\prime\prime\prime}$. 
	
	At the point 3.2.2, while solving task 2 it had been concluded that the coordinates of events in the systems $S$ and $S^\prime$ are related to any system of equations (2.2) - (2.5), and in the systems $S$ and $S^{\prime\prime\prime}$ - with $(2.2_{\slin{\nkg L}}) - (2.5_{\slin{\nkg L}})$, with the substitution of $v$ into $-v$ and $^\prime$ into $^{\prime\prime\prime}$. Now it is convenient to make use of the systems (2.3) and $(2.3_{\slin{\nkg L}})$. The latter hadn't been put down in details before, so we give it here under the number (4.4):
		$$
\left\{
\begin{array}{l}
x - x_{\slin{\nkg L}}  =  \dfrac {(x^{\prime\prime\prime} - x^{\prime\prime\prime}_{\slin{\nkg L}})}{\gamma} - v(t - t_{\slin{\nkg L}}) \\ \\
t^{\prime\prime\prime} - t^{\prime\prime\prime}_{\slin{\nkg L}}  = \dfrac {t - t_{\slin{\nkg L}}}{\gamma}  + \dfrac{v}{c^2}(x^{\prime\prime\prime} - x^{\prime\prime\prime}_{\slin{\nkg L}}) \;.
\end{array}
\right. \eqno(4.4)
$$

	The change of the movement of the watch $C$ direction relatively to the axis $x^\prime$ took place at the point $x^\prime_{\slin{\nkg L}}$ of the axis $x^\prime$ at the moment of time  $t^\prime_{\slin{\nkg L}}$  by the watch $C^\prime_1$, fixed in this point (the event $\nkg L$). Hereby the watch $C^\prime_1$ was included in the IRS  $S^{\prime\prime\prime}$, preserving its coordinate on the axis $x^\prime$ (now  $x^{\prime\prime\prime}$) and the continuity of readings (see $(3.11_2)$ and (3.12)).
	
	The remaining watch of the system  $S^{\prime\prime\prime}$ also keeps its coordinate (see Note $^7$), but must be synchronized  according to $C^\prime_1$ (now $C^{\prime\prime\prime}_1$) transiting to the system  $S^{\prime\prime\prime}$.
	
	Let's study if it keeps the continuity of readings. For this purpose let's consider the watch $C^\prime_{x^\prime}$ of the system $S^\prime$, fixed at an arbitrary point $x^\prime$ of the axis $x^\prime$. At a certain moment the direction of the movement of the axis $x$  will change relatively to the point $x^\prime$, and the watch $C^\prime_{x^\prime}$ will be included into the system $S^{\prime\prime\prime}$. Let $t$ be the time of this transition by the watch of the system $S$, situated in the point $x$, coinciding with $x^\prime$ at this moment. Then the time $t^\prime$ by the watch $C^\prime_{x^\prime}$ in the system $S^\prime$ is calculated at the moment of this transition with the help of the system of equations (2.3), and the time $t^{\prime\prime\prime}$ at the same moment and by the same watch, but in the system $S^{\prime\prime\prime}$ is calculated using the system of equations (4.4).
	
	Subtracting termwise the equations $(2.3_2)$ from $(4.4_2)$  , taking into consideration the equalities $x^\prime = x^{\prime\prime\prime}$, (3.11), (3.12) and (1.1), we arrive at:
$$t^{\prime\prime\prime} - t^\prime = t^{\prime\prime\prime}_{\slin{\nkg L}} - \dfrac{t_{\slin{\nkg L}}}{\gamma} + \dfrac{v}{c^2}(x^{\prime\prime\prime}   + x^\prime - x^{\prime\prime\prime}_{\slin{\nkg L}}) =  t^\prime_{\slin{\nkg L}}\left(1 - \dfrac{1}{\gamma^2} + \dfrac{v^2}{c^2}\right) + \dfrac{v}{c^2}2x^\prime = $$ $$ = \dfrac{2v^2}{c^2}t^\prime_{\slin{\nkg L}} +  \dfrac{2v}{c^2}x^\prime =  \dfrac{2v}{c^2} (x^\prime - x^\prime_{\slin{\nkg L}}).$$

	So the readings of the watch in the system $S^\prime$, situated in the point $x^\prime$ (i.e. $C^\prime_{x^\prime}$) at the moment of getting included in the system $S^{\prime\prime\prime}$ undergo a leap
	\begin{center}   
   $\qquad\qquad \textit{The leap} $ = $t^{\prime\prime\prime} - t^\prime =  \dfrac{2v}{c^2}(x^\prime - x^\prime_{\slin{\nkg L}}).\qquad\qquad\qquad\qquad\qquad\qquad (4.5)$
 \end{center}
 
	Considering (3.11) and (3.15) in particular, the readings of the watch $C^\prime$, situated in the point $x^\prime = 0$, undergo a leap 
		 $$t^{\prime\prime\prime}_{\slin{\nkg R}} - t^\prime_{\slin{\nkg R}} = \dfrac{2\gamma v^2}{c^2}t_{\slin{\nkg R}}\;,\eqno (4.6)$$
while getting included into the system $S^{\prime\prime\prime}$, i.e. at the moment of the event $\nkg R$,
hence its readings should be increased by this quantity due to the resynchronization.

	Now it's clear that at the moment of the second meeting the readings of the watch ${C^\prime}$ will be greater than the time read by it between the first and the second meetings as much as its readings were increased with the leap at the moment of the event $\nkg R$ between the first and the second meetings, i.e.	$\dfrac{2\gamma v^2}{c^2}t_{\slin{\nkg R}}$  greater.
	
	More formally, the time read by the watch ${C^\prime}$ (later renamed into $C^{\prime\prime\prime}$) between the events $\nkg B$ and $\nkg E$, taking into consideration (4.6), (3.16) and (1.1) makes
	$$\tau^{\prime\prime\prime}_{\slin{\nkg E}} = (t^{\prime\prime\prime}_{\slin{\nkg E}} - t^{\prime\prime\prime}_{\slin{\nkg R}}) + (t^\prime_{\slin{\nkg R}} - t^\prime_{\slin{\nkg B}}) = t^{\prime\prime\prime}_{\slin{\nkg E}} - (t^{\prime\prime\prime}_{\slin{\nkg R}} - t^\prime_{\slin{\nkg R}}) = t^{\prime\prime\prime}_{\slin{\nkg E}} - \dfrac{2\gamma v^2}{c^2}t_{\slin{\nkg R}} = \dfrac{2t_{\slin{\nkg R}}}{\gamma}.$$
	So in task 2 the time read by the watches $C$ and $C^\prime$ from the first to the second meetings makes accordingly
	$$\tau_{\slin{\nkg E}} = 2t_{\slin{\nkg R}}\;,  \tau^{\prime\prime\prime}_{\slin{\nkg E}} = 2t_{\slin{\nkg R}}/\gamma\;, \eqno (4.7)$$ 
That coincides with the results (4.3) got in task 1.

\textit{	So the paradox disappears, and there is only a misunderstanding,  because the readings of the watch ${C^\prime}$ at the moment of its second meeting in task 2 were identified with the time read by it between the first and the second meetings.}

	\textbf{4.3} \quad  To clear everything up, let's study the relation between the systems $S^{\prime\prime}$ and $S^{\prime\prime\prime}$. In each of them the axis $x^\prime$ after the turn serves as the axis of spatial coordinates, in one system - under the name of $x^{\prime\prime}$, in the other - as $x^{\prime\prime\prime}$. So these systems are static as to each other. 
	
	It is convenient to consider as the systems, relating $S^{\prime\prime}$ and $S^{\prime\prime\prime}$ with $S$ the systems obtained  from (2.1) through the change of $v$ into $-v$, $\nkg A$ and symbol $^\prime$ into $\nkg R$ and symbol $^{\prime\prime}$ in the first case, and into $\nkg L$ and symbol $^{\prime\prime\prime}$ - in the second case. Let's write out these systems under the numbers (4.8), (4.9) accordingly.
						$$
\left\{
\begin{array}{l}
x^{\prime\prime} - x^{\prime\prime}_{\slin {\nkg R}} = \gamma \left\{(x - x_{\slin {\nkg R}}) + v(t - t_{\slin {\nkg R}})\right\}\\
t^{\prime\prime} - t^{\prime\prime}_{\slin {\nkg R}} = \gamma \left\{(t - t_{\slin {\nkg R}}) + \dfrac{v}{c^2}(x - x_{\slin {\nkg R}})\right\},
\end{array}
\right. \eqno(4.8)
$$
$$
\left\{
\begin{array}{l}
x^{\prime\prime\prime} - x^{\prime\prime\prime}_{\slin {\nkg L}} = \gamma \left\{(x - x_{\slin {\nkg L}}) + v(t - t_{\slin {\nkg L}})\right\}\\
t^{\prime\prime\prime} - t^{\prime\prime\prime}_{\slin {\nkg L}} = \gamma \left\{(t - t_{\slin {\nkg L}}) + \dfrac{v}{c^2}(x - x_{\slin {\nkg L}})\right\}.
\end{array}
\right. \eqno(4.9)
$$ 
	Subtracting  termwise the equations $(4.8_1)$ from ($(4.9_1)$ and $(4.8_2)$) from $(4.9_2)$, we get
$$
\left\{
\begin{array}{l}
x^{\prime\prime\prime} - x^{\prime\prime} = (x^{\prime\prime\prime}_{\slin {\nkg L}} - x^{\prime\prime}_{\slin {\nkg R}}) + \gamma\{(x_{\slin {\nkg R}} - x_{\slin {\nkg L}}) + v(t_{\slin {\nkg R}} - t_{\slin {\nkg L}})\}\\
t^{\prime\prime\prime} - t^{\prime\prime} = (t^{\prime\prime\prime}_{\slin {\nkg L}} - t^{\prime\prime}_{\slin {\nkg R}}) + \gamma \left\{(t_{\slin {\nkg R}} - t_{\slin {\nkg L}}) + \dfrac{v}{c^2}(x_{\slin {\nkg R}} - x_{\slin {\nkg L}})\right\}.
\end{array}
\right. 
$$	According to $(3.5_1)$, $(3.11_1)$ and (3.15)
		$$x_{\slin {\nkg R}} - x_{\slin {\nkg L}} = vt_{\slin {\nkg R}}\; ,\; t_{\slin {\nkg R}} - t_{\slin {\nkg L}} = 0 \; ; $$
and according to (3.6), (3.12), (3.15) and (1.1)
$$x^{\prime\prime\prime}_{\slin {\nkg L}} - x^{\prime\prime}_{\slin {\nkg R}} = -v\gamma t_{\slin {\nkg R}}\; ,\; t^{\prime\prime\prime}_{\slin {\nkg L}} - t^{\prime\prime}_{\slin {\nkg R}} = \gamma t_{\slin {\nkg R}}(1 - 1/\gamma^2) = \dfrac{\gamma v^2}{c^2}t_{\slin {\nkg R}}\;.$$
	Substituting it into the last system of equations, we get
		$$
\left\{
\begin{array}{l} 
x^{\prime\prime\prime} - x^{\prime\prime} = 0\\
t^{\prime\prime\prime} - t^{\prime\prime} = \dfrac{2\gamma v^2}{c^2}t_{\slin {\nkg R}}.
\end{array}
\right. \eqno (4.10)
$$

Hence, the IRS $S^{\prime\prime\prime}$ and $S^{\prime\prime}$, which in literature before were considered as coinciding , have one coordinate axis in common, but differ in the shift of watch readings in the system $S^{\prime\prime\prime}$ by the quantity $\dfrac{2\gamma v^2}{c^2}t_{\slin {\nkg R}}$  comparing with the watch of the system $S^{\prime\prime}$, situated in the same point of the coordinate axis.

\section{	Conclusion}

Let's describe the gist of the above-mentioned things in short.

1.	The axis $x$ is included into the IRS $S$ all the time. The axis $x^\prime$ is at first included into the IRS $S^\prime$ and moves along the axis $x$ in the positive direction (to the right). Because after some time it changes its direction of movement relatively to the IRS $S$, it cannot remain in the IRS $S^\prime$ and is included into another one.

2.	It turned out that in transition of the axis $x^\prime$ from $S^\prime$ to another IRS, only one watch  of this axis can keep the continuity of readings; the readings of the other one  by the leap should be changed by the quantity, depending on its position on the axis $x^\prime$. These changes weren't realized, but they were automatically taken into consideration in Lorenz's transformations. 

3.	Solving tasks 1 and 2 about the watches'  readings at the moment of their second meeting (the moment of watches' return to each other), it was always admitted without special reservations, consciously or not that the continuity of readings in the first task was kept by the watch $C^\prime$, situated in the origin of coordinates of the axis $x^\prime$ and in the second task - by the watch $C^\prime_1$ situated on the axis $x^\prime$ in the point of the utmost left position of the watch $C$ on this axis.

	That is why in the suppositions of  task 2 the readings of the watch $C^\prime$ are increased with the leap by such a quantity that by the moment of  return, its readings will be by so many times greater than the readings of the watch $C$, as they were less in the suppositions of task 1.

4.	These correct results were considered before as mutually exclusive (the watch paradox) because tasks 1 and 2 were considered in essence before as one and the same, but  variously formulated task. Meanwhile, with unspoken suppositions, these are different tasks, because in the first one the continuity of readings is kept with the watch $C^\prime$, and in the second -  with $C^\prime_1$; and as a result, the readings of $C^\prime$ are here compulsorily increased with the leap.

Because this paradox caused doubts in the inner consistency of SR, many even great specialists (see Introduction) "preferred" to doubt in the correctness of the solving of task 2 than to agree with its results. The solving of task 1 within the framework of SR caused no doubts.

5.	After it was shown in section 4 that considering the leap of watch readings, the time read by the watch $C^\prime$ at the moment of the return, is the same in tasks 1 and 2; the  "great" watch paradox, threatening the inner consistency of SR, was solved.

However, "the small" watch paradox still remains unsolved.  It's quite unclear why the watch $C^\prime$  should return with the readings less than the readings of the watch $C$. Because at the sections of even movement  the watch $C^\prime$, according to SR, is just in the same conditions as the watch C, and the fact of changing the direction cannot, according to Laue, make much difference, the sufficient length of these sections provided.\\ 

I gratefully thank  Lyudmila Timoshenko  Ph.D., Department of Foreign Languages at the Bashkir State University, for translation of the text of the manuscript.

\begin{center}\large{REFERENCES}\end{center}

\noindent 1. A. Einstein, Zur Elertrodynamik der bewegter K\"{o}rper. Ann. Phys., 1905.\\
2. P. Langevin, L' \'evolution de l\'espace et du temps. Scientia, 10, 31, 1911.\\
3. M. von Laue, Das Relativt\"{a}tsprinzip. Jahrb\"{u}cher der Philosophie,

Berlin, 1913.\\
4. M. von Laue, Das Relativt\"{a}tsprinzip. Braunschweig, 1913.\\
5. L. Marder, Time and space - traveler. London, 1971.\\
6. I.I. Goldenblat,   Time paradoxes in relativity mechanics. Nauka, Moscow,

 1972 (in Russian).\\
7. T. Grandou, J.L. Rubin, Twin Paradox and Causality.

 arXiv:0704.2736v1 [gr-gc] 20Apr2007.\\ 
8. A. Einstein, Eine Dialog \"{u}ber Einw\"{a}nde gegen die Relativt\"{a}tstheorie.

Naturwiss., 6, 1918.  \\
9. W. Pauli, Theory of relativity . Pergamon Press, 1958.\\
10. V.A. Fok,   Theory of space, time and gravity. GITTL, Moscow, 1955

(in Russian). \\
11. Ya.P. Terletskiy,   Paradoxes of theory of relativity. Nauka, Moscow 1966 

(in Russian).\\
12. D. Bom , The special theory of relativity. New  York -- Amsterdam.1965.\\
13. M. Born, Einstein's theory of relativity. New York, 1962.\\
14. L.I. Mandelshtam,   Lectures on optic, theory of relativity and quantum 

 mechanics. Nauka, Moscow 1972 (in Russian).\\
15. C. M{\o}ller, The  theory of relativity. Oxford, 1972.\\
16. L.D. Landau  and E.M. Lifshits. 1988. Theoretical physics, Vol. 2, Field 

theory . Nauka, Moscow (in Russian).\\
17. R.P. Feinman, R.B. Leighton, M. Sands, The Feinman lectures on

physics, Mainly mechanics, radiation, and heat. London, 1977.\\
18. M. Born, Ein Besuch bei den Raumfahrern und das Uhrenparadoxon.

Physik,~ Bl. 14, 1958.\\
19. E.F. Taylor, J.A. Wheeler, Spacetime Physics. San Francisco, 1966.\\
20. H. Boundi, Assumption and myth in physical theory, Cambridge, 1967.\\
21. L. Brillouin, Relativity reexamined. New York -- London, 1970.\\
22. A.F. Kracklauer, P.T. Kracklauer, On the Twin Non - paradox.

 arXiv:physics/0012041v1 [physics.gen-ph] 18Dec2000.\\
23. A. Einstein, Principe de relativit\'{e} et ses cons\'{e}quences dans la physique 

moderne. Arch. sci. phys. Natur., ser.4, 29, 1910. \\ 

\end{document}